\def\year{2021}\relax
\documentclass[letterpaper]{article} 
\usepackage{aaai21}  
\usepackage{times}  
\usepackage{helvet} 
\usepackage{courier}  
\usepackage[hyphens]{url}  
\usepackage{graphicx} 
\usepackage{natbib}  
\usepackage{caption} 
\title{A Deep Learning Algorithm for Piecewise Linear Interface Construction (PLIC)}
\author{
    Mohammadmehdi Ataei,\textsuperscript{\rm 1,2} 
    Erfan Pirmorad,\textsuperscript{\rm 2} 
    Franco Costa,\textsuperscript{\rm 3} 
    Sejin Han,\textsuperscript{\rm 4} 
    Chul B Park,\textsuperscript{\rm 2}
    Markus Bussmann\textsuperscript{\rm 2} \\ 
}
\affiliations{
    \textsuperscript{\rm 1}Vector Institute for Artificial Intelligence, 661 University Ave Suite 710, Toronto, ON M5G 1M1, Canada\\
    \textsuperscript{\rm 2}University of Toronto, 5 King's College Rd, Toronto, ON M5S 3G8, Canada\\
    \textsuperscript{\rm 3}Autodesk, Inc., 259-261 Colchester Rd., Kilsyth, VIC. 3137, Australia\\
    \textsuperscript{\rm 4}Autodesk, Inc. 2353 North Triphammer Rd., Ithaca, NY 14850, USA\\

    \{ataei, pirmorad\}@mie.utoronto.ca, \{franco.costa, sejin.han\}@autodesk.com, \{park, bussmann\}@mie.utoronto.ca

}

\begin{document}

\maketitle

\begin{abstract}
	 Piecewise Linear Interface Construction (PLIC) is frequently used to geometrically reconstruct fluid interfaces in Computational Fluid Dynamics (CFD) modeling of two-phase flows. PLIC reconstructs interfaces from a scalar field that represents the volume fraction of each phase in each computational cell. Given the volume fraction and interface normal, the location of a linear interface is uniquely defined. For a cubic computational cell (3D), the position of the planar interface is determined by intersecting the cube with a plane, such that the volume of the resulting truncated polyhedron cell is equal to the volume fraction. Yet it is geometrically complex to find the exact position of the plane, and it involves calculations that can be a computational bottleneck of many CFD models. However, while the forward problem of 3D PLIC is challenging, the inverse problem, of finding the volume of the truncated polyhedron cell given a defined plane, is simple. In this work, we propose a deep learning model for the solution to the forward problem of PLIC by only making use of its inverse problem. The proposed model is up to several orders of magnitude faster than traditional schemes, which significantly reduces the computational bottleneck of PLIC in CFD simulations.
	
\end{abstract}
\section{Introduction}
\noindent 

\begin{figure}[ht]
	\centering
	\includegraphics[width=0.9\linewidth]{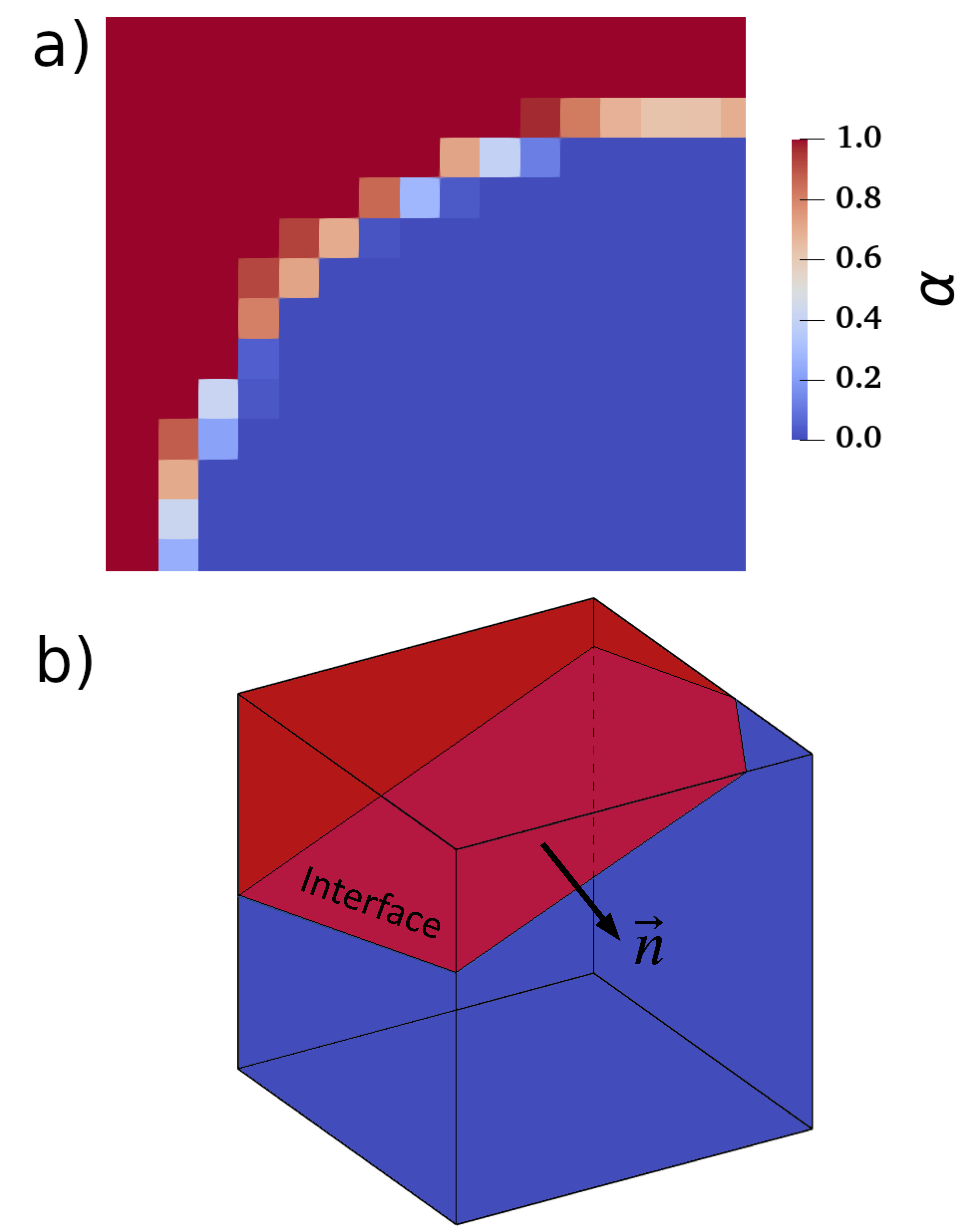}
	\caption{a) An example of a VOF scalar field for a liquid-gas system. $\alpha=1$ in liquid cells, $\alpha=0$ in gas cells, and $0 < \alpha < 1$ in interface cells. b) Interface reconstruction using PLIC.}
	\label{fig:PLIC}
\end{figure}

\begin{figure*}
	\centering
	\includegraphics[width=1\linewidth]{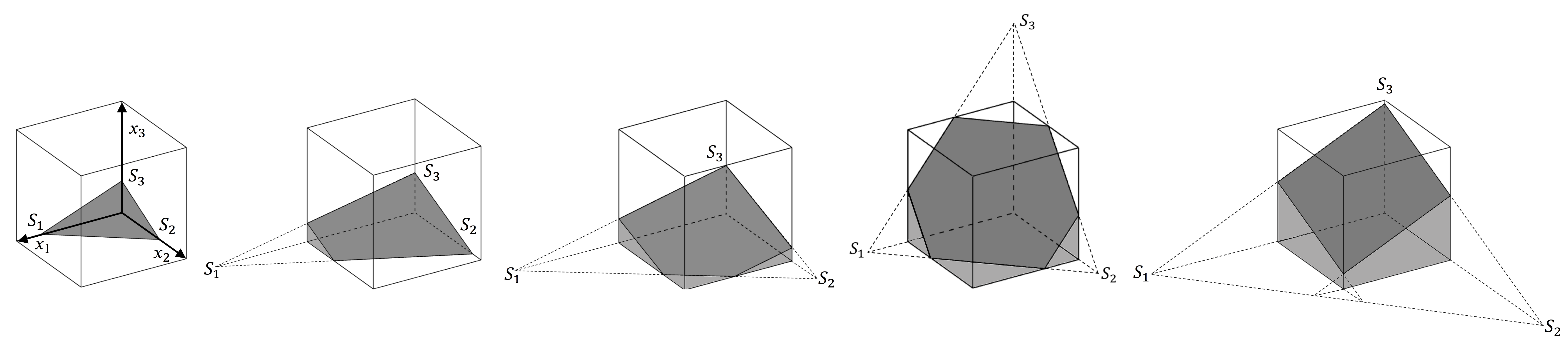}
	\caption{All possible cases for intersection of a plane and a unit cube.}
	\label{fig:intersections}
\end{figure*}

\begin{figure}[ht]
	\centering
	\includegraphics[width=\linewidth]{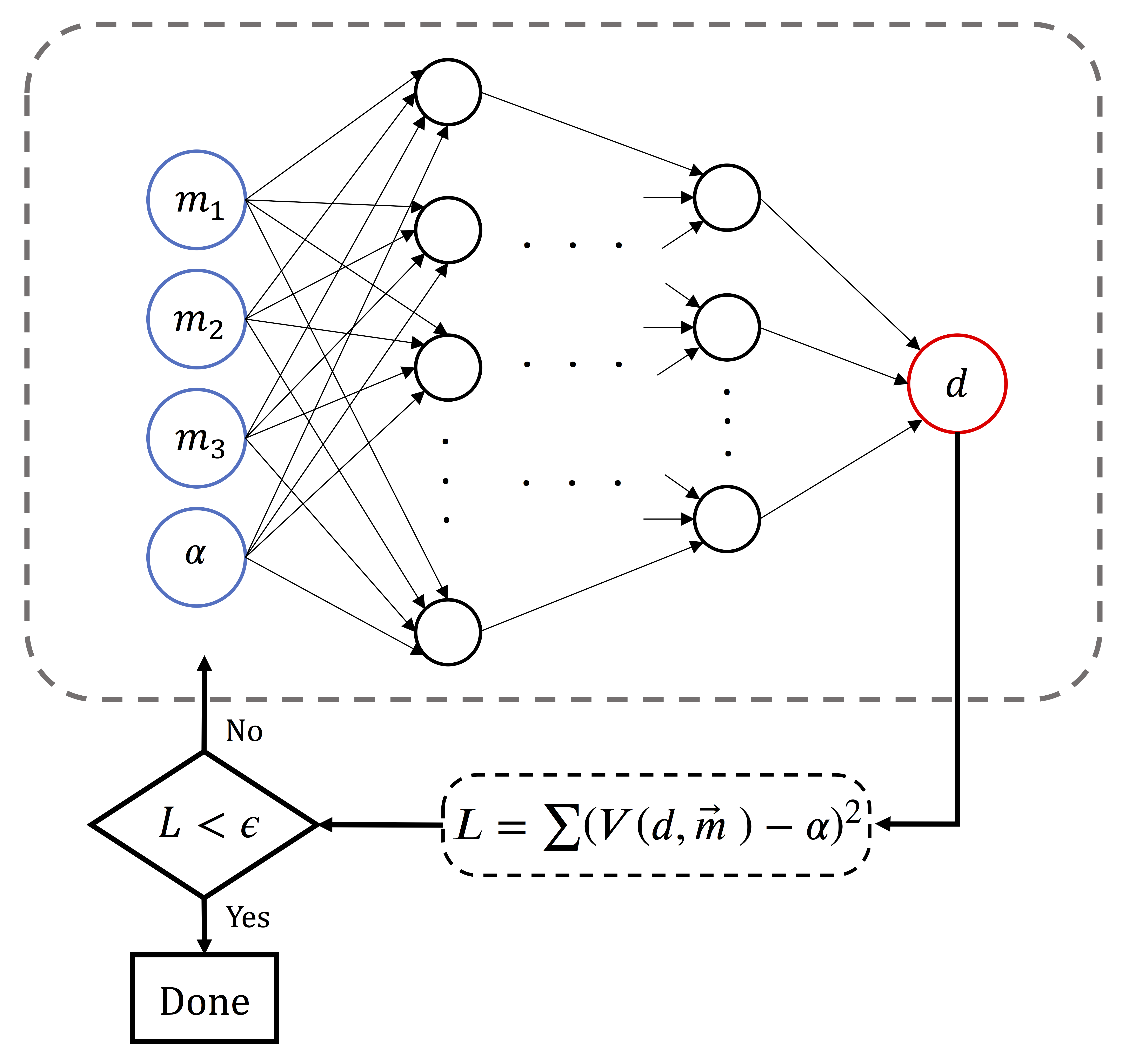}
	\caption{DNN architecture.}
	\label{fig:NN}
\end{figure}

Multiphase flows (flows consisting of more than one component) are ubiquitous in nature and are of interest in a variety of practical applications, and as such a significant amount of work has been conducted on the computational modeling of such flows (e.g.\ \citealt{Balachandar2020,Rothman1994,WANG2016}). One of the challenges of multiphase flow modeling is to track interfaces between different phases. Among numerous Eulerian and Lagrangian interface tracking schemes, the Volume-of-Fluid (VOF) Eulerian method is widely used \cite{HIRT1981}.

In the VOF method, as shown in figure \ref{fig:PLIC}a, the volume fraction of one the two phases within each computational cell is represented by a scalar field $\alpha$. For example, for a liquid-gas system, $\alpha=1$ or $\alpha=0$ in each computational cell that is occupied by the liquid or gas phase, respectively, and $0 < \alpha < 1$ in interface cells. An advection equation is solved for $\alpha$ to track the location of interface throughout the simulation. There are two major categories of VOF schemes: algebraic and geometric \cite{LAFAURIE1994}. Geometric VOF schemes provide more accurate fluid-fluid interface reconstruction compared to algebraic schemes, but require more complicated geometrical calculations. Of these methods, Piecewise Linear Interface Construction (PLIC) is frequently used to geometrically reconstruct an interface as a line (2D) or plane (3D) \cite{YOUNGS1982}. As shown in figure \ref{fig:PLIC}b, for a cubic cell, PLIC linearly approximates an interface, such that the volume of the resulting truncated polyhedron cell is equal to the volume fraction $\alpha$, where the plane normal unit vector $\vec{n}$ is calculated as the gradient of the volume fraction scalar field, as $ \vec{n} = - \frac{\nabla \alpha}{||\nabla \alpha||}$ \cite{normalCalculation}. Eliminating symmetrical cases, the number of possible cube-plane intersections can be reduced to the five scenarios shown in figure \ref{fig:intersections}, where $\alpha \leq 0.5$ (for $\alpha > 0.5$, $\alpha$ is changed to $1-\alpha$ to match one of the five cases). The constant of the plane $d$ is the smallest distance from the plane to the origin of the Cartesian coordinate that conforms with the cubic cell edges, which must be calculated. The problem of computing $d$ is known as the forward problem of PLIC \cite{SCARDOVELLI2000228}. 

To date, many studies have used PLIC for interface reconstruction on structured and unstructured grids, and also for related calculations such as calculating the distance between two interfaces, and have proposed improvements to reduce the implementation complexity and increase its efficiency \cite{LOPEZ200551, ATAEI2021107698, Dai2019, ataei2021nplic, Nahed2020, SCHEUFLER2019}. A simple analytical solution can be used to solve PLIC in 2D for a square cell. However, analytical solutions in 3D for a cubic cell include multiple \textit{if-else} conditions, and require more sophisticated operations (e.g.\ dealing with the inverse of third degree polynomials and complex numbers) \cite{SCARDOVELLI2000228, YANG200641}. For example, Scardovelli and Zaleski \cite{SCARDOVELLI2000228} found an analytical solution to the forward problem by inverting the inverse problem of PLIC (i.e.\ finding the volume fraction of a polyhedron formed given a plane with $\vec{n}$ and $d$), which is a simpler geometric calculation. This analytical solution includes a number of conditional statements to select a solution from a set of equations, corresponding to different cases of the intersection of a plane with a cube. This selection can slow down a CFD solver, and is not easily parallelizable. To reduce the computational cost, alternative iterative and approximation techniques have been developed to calculate $d$ \cite{iterative, iterative2, KAWANO2016130}.

Machine Learning (ML) has attracted considerable attention in the computational sciences, including CFD. Over the last decade, ML techniques have been used extensively in fluid mechanics to solve a variety of problems, including model discovery, modeling of reduced order, processing of experimental data, shape optimization, and flow control \cite{annurev-fluid2019, Raissi2020,Brenner2019, Duraisamy2017,pfaff2020learning,Nikolaev2020DeepLF,DElia2020,Yoon2020}. Previously, we have published an alternative neural network based technique to perform PLIC calculations, by training a neural network with a large synthetic dataset of PLIC solutions \cite{ataei2021nplic}. In this work, we propose a direct solution to the forward problem of PLIC using a fully connected Deep Neural Network (DNN). Unlike the previous work, we demonstrate that a DNN can solve the forward problem given only the inverse problem of PLIC, without requiring a synthetic dataset. We show that, in addition to its simplicity, this approach significantly outperforms other traditional PLIC implementations.

\section{Methodology}
\label{Methodology}

We solve the forward problem of PLIC by converting it into an optimization problem that can be solved using a DNN. In the forward problem, we must find $d$ such that the volume $V$ of a polyhedron cell is equal to the volume fraction $\alpha$ in a unit cubic cell:

\begin{equation}
	V = \alpha
\end{equation}

\noindent This equation can be converted into an optimization problem for a DNN by defining a loss function $L$ as:

\begin{equation}
	L = \sum_{batch} (V(d,\vec{m}) - \alpha)^2
	\label{eqn:loss}
\end{equation}

\noindent where $\vec{m} = (m_1,m_2,m_3)$ are the components of $\vec{n}$ along the axes shown in figure \ref{fig:intersections} \cite{Ashgriz}, calculated as:
\begin{equation}
    \begin{array}{l}
    m_{1}:=\min \left(\left|n_{x}\right|,\left|n_{y}\right|,\left|n_{z}\right|\right) \geq 0 \\
    m_{3}:=\max \left(\left|n_{x}\right|,\left|n_{y}\right|,\left|n_{z}\right|\right)>0 \\
    m_{2}:= \left|n_{x}\right| + \left|n_{y}\right| + \left|n_{z}\right| - m_{1} - m_{3} \geq 0
    \end{array}
\end{equation}

As shown in figure \ref{fig:NN}, we design a multilayer perceptron (MLP) neural network with fully connected layers, comprised of an input layer, $H$ hidden layers each with $N$ neurons, and an output layer, to solve this optimization problem. The network outputs $d$ based on $\vec{m}$ and $\alpha$ as inputs. If the difference between $V(d,\vec{m})$, calculated based on the predicted $d$ (the output of the DNN), and the known $\alpha$ is sufficiently minimized, then the forward PLIC problem is solved by the DNN. One of the advantages of this technique is that (unlike conventional methods for training a DNN), there is no need for synthetic data to train the DNN, as the DNN automatically finds the solution to the forward PLIC problem by minimizing this loss function.

Fortunately, formulating $V$ (i.e.\ the inverse problem of PLIC) in terms of the plane parameters $(d,\vec{m})$ is a straightforward geometrical problem. The volume under the plane can be found by calculating the volume of the pyramid formed between the intersection of the plane with the three coordinates ($S_{1},S_{2},S_{3}$), and then subtracting the volume of the parts that extend beyond the cube (see figure \ref{fig:intersections}). Such calculations result in a set of equations for $V$ for each of the five cases, as follows \cite{SCARDOVELLI2000228}:

\begin{equation}
	\footnotesize
	V = \frac{1}{\lambda} \cdot\left\{\begin{array}{ll}
	f_1 (d,\vec{m}) &  0 \leq d \leq m_1 \\
	f_2 (d,\vec{m}) &  m_1 \leq d \leq m_2 \\
	f_3 (d,\vec{m}) &  m_2 \leq d \leq \min \left(m_1+m_2, m_3\right) \\
	f_4 (d,\vec{m}) &  m_3 \leq d \\
	f_5 (d,\vec{m}) &  \min \left(m_1+m_2, m_3\right) \leq d \leq m_3
	\end{array}\right.
	\label{eqn:conditions}
\end{equation}

\noindent where:

\begin{equation}
	\begin{array}{ll}
	f_1 &= d^{3} \\
	f_2 &= d^{3}-\delta_1^{3}  \\
	f_3 &= d^{3}-\delta_1^{3}-\delta_2^{3} \\
	f_4 &= d^{3}-\delta_1^{3}-\delta_2^{3}-\delta_3^{3}  \\
	f_5 &= 6 m_1 m_2\left(d-\frac{1}{2}\left(m_1+m_2\right)\right) 
	\end{array}
\end{equation}

\noindent where:

\begin{equation}
	\begin{array}{ll}
		\delta_1 & = d-m_1               \\
		\delta_2 & = d-m_2               \\
		\delta_3 & = d-m_3               \\
		\lambda  & = 6 \ m_1 m_2 m_3 
	\end{array}
\end{equation}

This DNN is implemented in the Pytorch deep learning library \cite{NEURIPS2019_bdbca288}. We use ReLU activation functions for all layers except for the output layer, which is a linear activation function. The gradient of the custom loss function in Eq.\ \ref{eqn:loss} is calculated using the Pytorch automatic differentiation \textit{autograd} package.

For the inputs of the DNN $\vec{m}$ is uniformly sampled where all three coordinates are positive. Similarly, $\alpha$ is sampled uniformly between $0$ and $0.5$. We used 5000 randomly shuffled sample points in total, split into three datasets: $70\%$ training, $20\%$ test, and $10\%$ validation. The model was trained on the training dataset, the validation dataset was used to ensure that the model was not overfitting during the training, and the test dataset was used to evaluate the final accuracy of the model.

The network weights and biases were randomly initialized at the beginning using Xavier initialization \cite{Xavier2010}. The results presented here are the average of six runs with six different random seeds for network initialization. The weights and biases were tuned using the Adam Optimizer \cite{Adam} with a batch size of 64 for 512 epochs.

\section{Results}

\subsection{Evaluation metrics}

The training was performed on a PC running Ubuntu 20.04 (Intel i7-8700K, NVIDIA 1080 Ti graphics card (CUDA version 11.0), 32GBs of RAM). It takes about 20 minutes to train each network. 

\begin{figure}[ht]
	\centering
	\includegraphics[width=\linewidth]{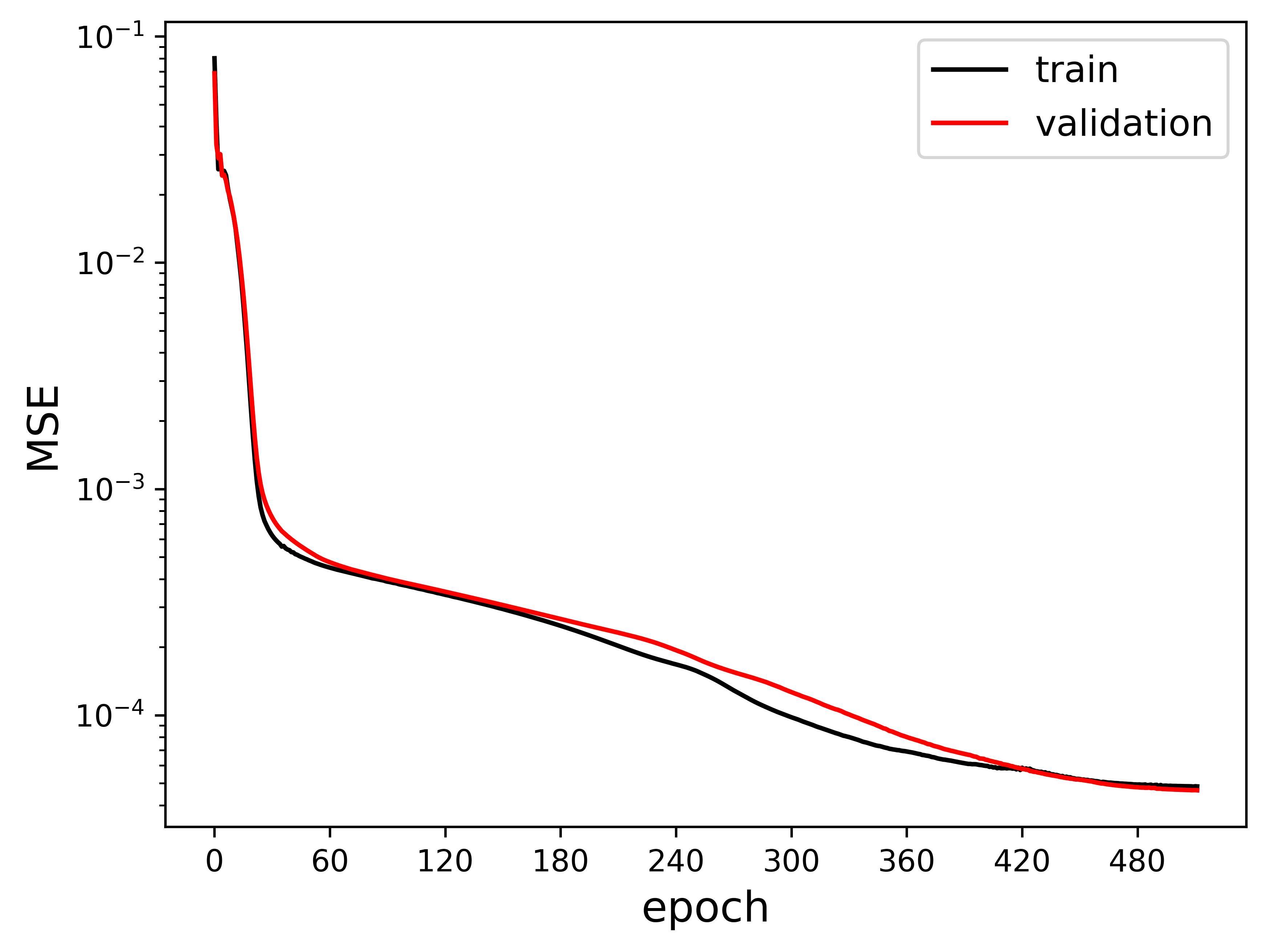}
	\caption{Training progress over 512 epochs with batch size of 64 for a DNN with one hidden layer with 24 neurons with seed = 100.}
	\label{fig:training}
\end{figure}

\begin{figure}[ht]
	\centering
	\includegraphics[width=\linewidth]{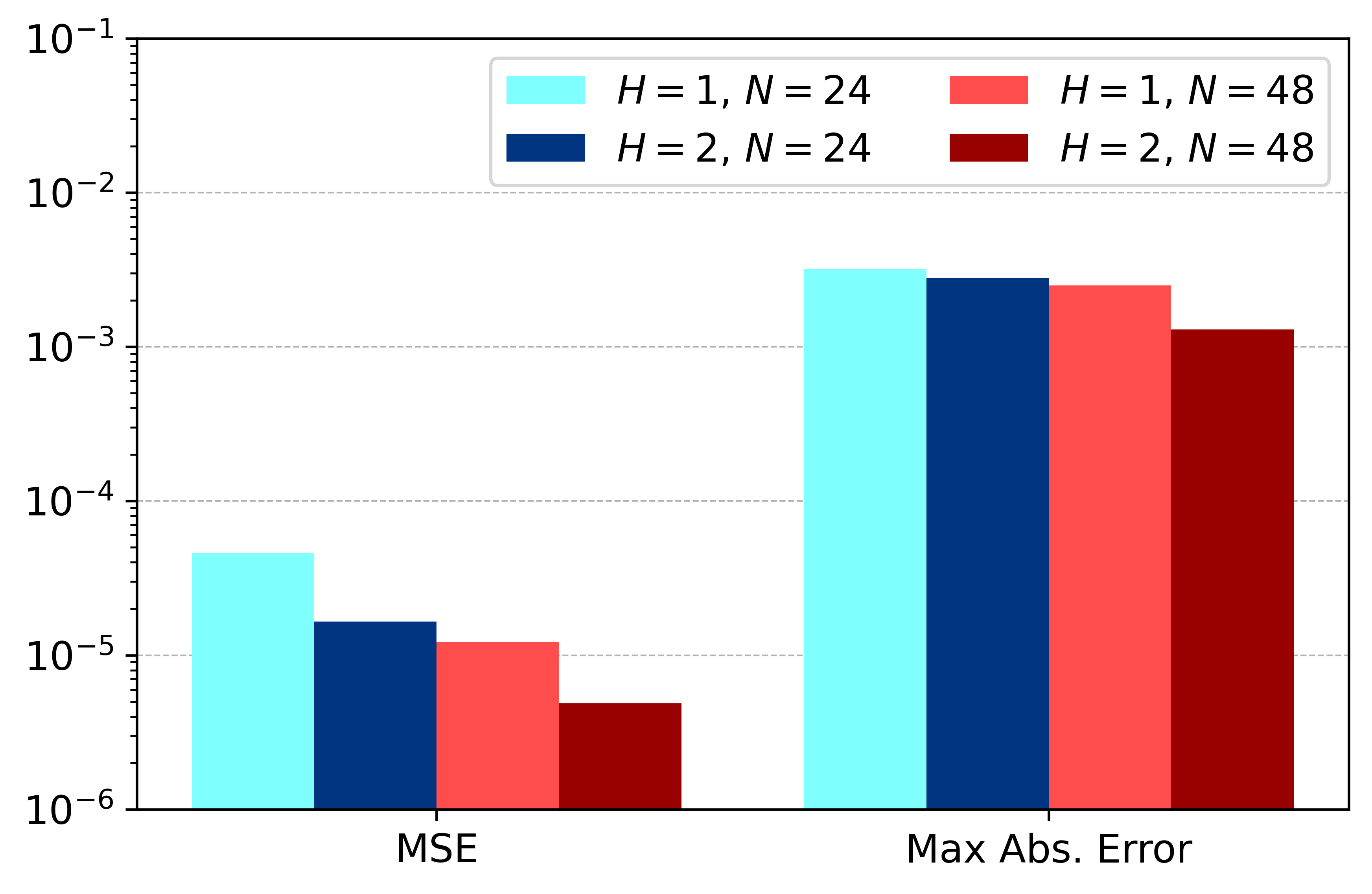}
	\caption{Mean Squared Error (MSE) and maximum absolute error on the test dataset, for different DNN models with different $H$ and $N$.}
	\label{fig:metrics}
\end{figure}

The training progress of a DNN with $H=1,\ N=24$ is shown in figure \ref{fig:training}. The Mean Squared Error (MSE) of both training and validation datasets decreased during training. The model performance was evaluated against the test dataset. Figure \ref{fig:metrics} shows the MSE and the maximum absolute error for DNNs with different $H$ and $N$. Notice the network with $H=2,\ N=48$ is the most accurate; however, other networks with fewer neurons and hidden layers also perform on-par or better when compared to PLIC models that approximate PLIC  (e.g.\
\citealt{KAWANO2016130}).

\begin{figure} [ht]
	\centering
	\includegraphics[width=\linewidth]{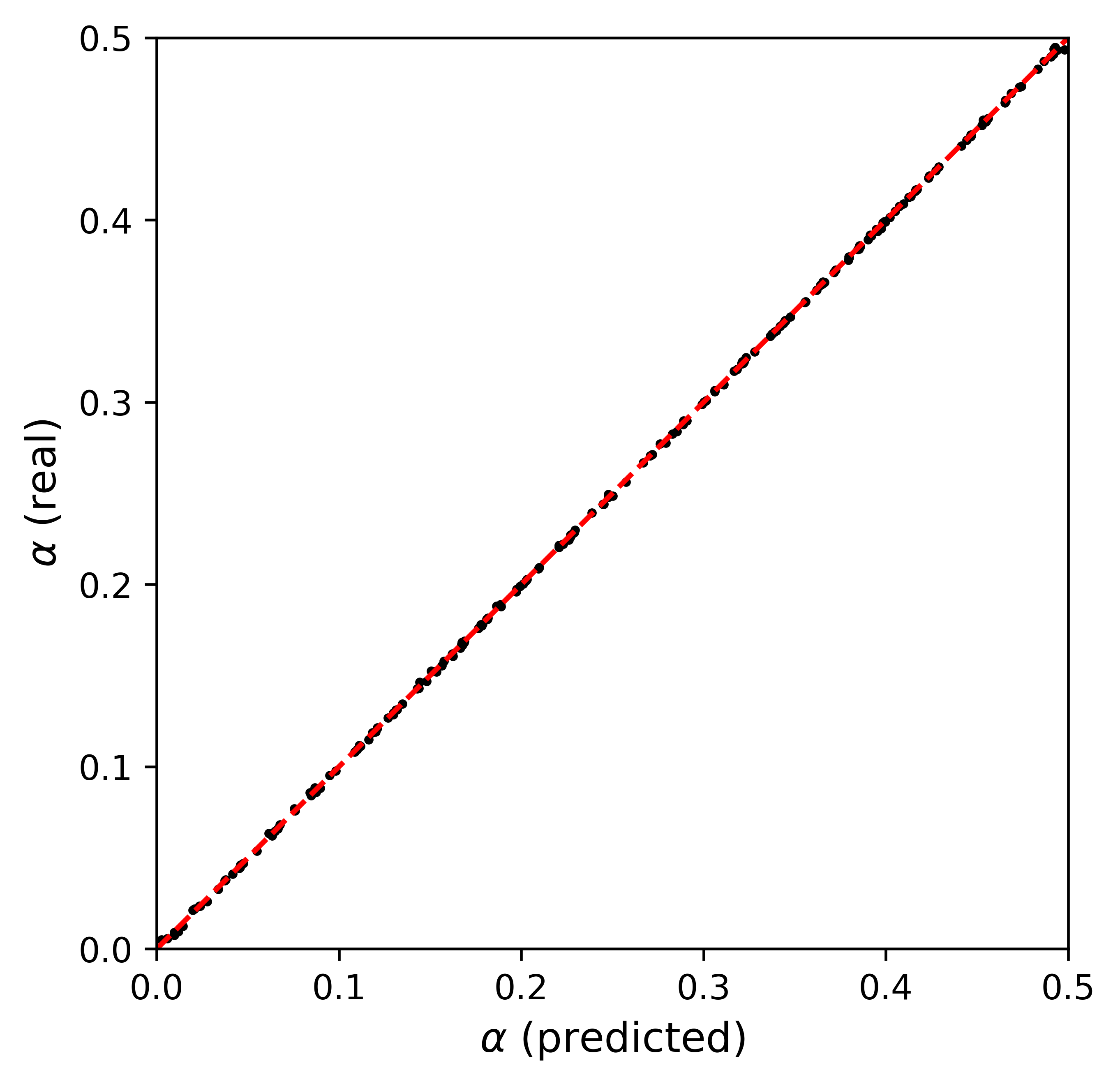}
	\caption{Plots of the DNN predictions vs real $\alpha$ by a DNN with two hidden layers each with 48 neurons (seed = 400). To avoid clutter, only 200 random points are selected from the test dataset.}
	\label{fig:NPLIC48}
\end{figure}

To visualize the accuracy of the DNN model, figure \ref{fig:NPLIC48} plots the value of $\alpha$ based on the value of $d$ predicted by the DNN ($H=2,\ N=48$) against the real value of $\alpha$, for 200 random data points from the test dataset. It can be seen that there is excellent agreement between the predicted $\alpha$ and real $\alpha$, close to the ideal where all points would lie on the diagonal line.

\subsection{Speedup}

The advantages of using a DNN for PLIC calculations is that it performs much faster than an analytical solution. Moreover, the DNN model can be easily parallelized on both CPUs and GPUs.

\begin{figure}[ht]
	\centering
	\includegraphics[width=\linewidth]{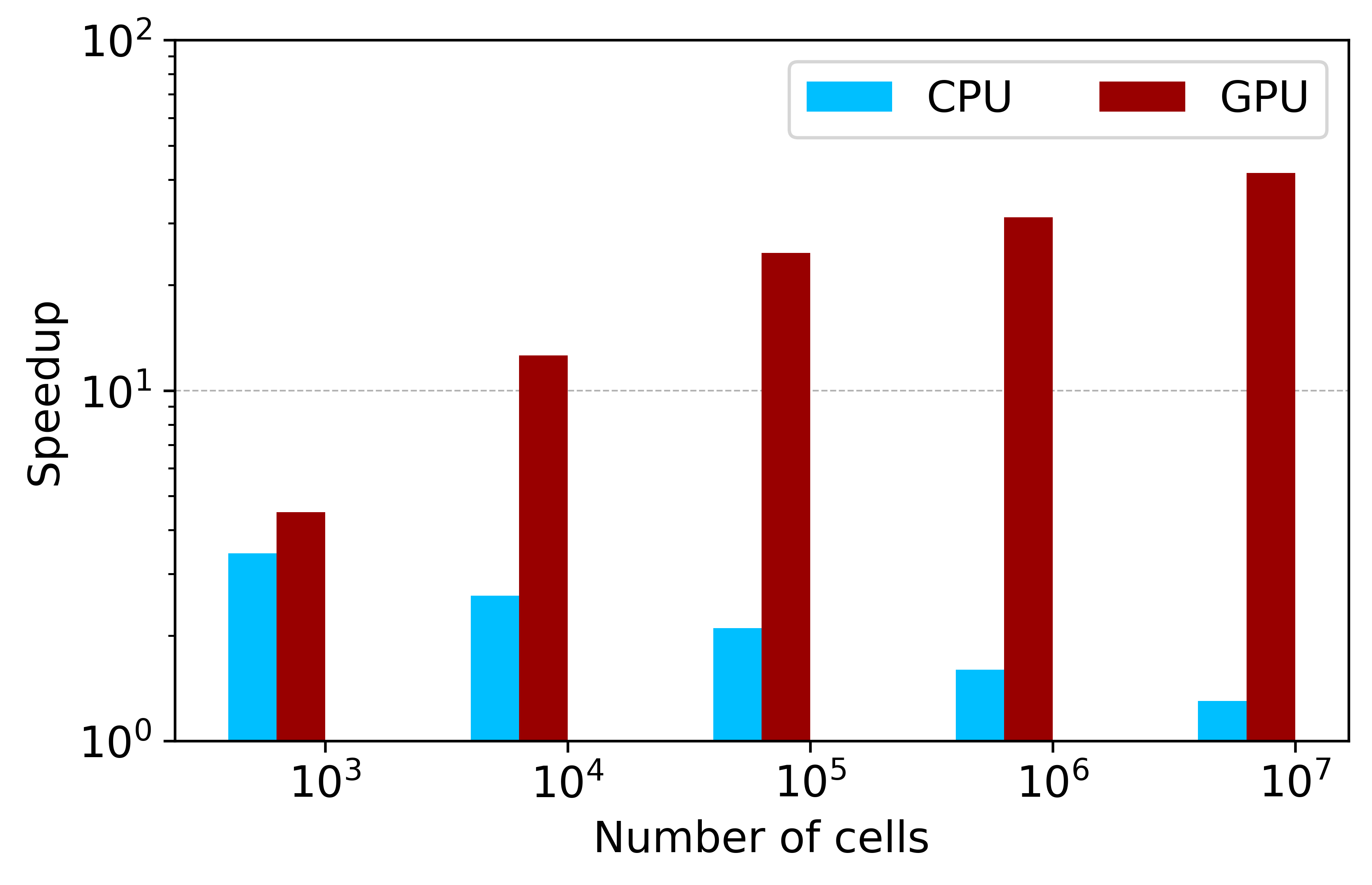}
	\caption{Speedups of the DNN model compared to the analytical solution of Scardovelli and Zaleski (running on one CPU core as the baseline).}
	\label{fig:speedups}
\end{figure}

Figure \ref{fig:speedups} shows a comparison of the DNN ($H=2$, $N=48$) versus the Scardovelli and Zaleski \cite{SCARDOVELLI2000228} analytical solution. To compare the models, the analytical solution was implemented in C++, and compiled using optimization flag {\tt -O3} using the GCC compiler version 9.3. The Pytorch DNN model was converted into Torchscript and imported into C++ as well. The \texttt{std::chrono} library and Pytorch's Profiler were used to measure the execution time of the Scardovelli and Zaleski PLIC solution and the neural model, respectively.

First, we compare the execution time of both models on one CPU core. Since the DNN can perform PLIC calculations in batches, we ran the models for different numbers of cells with randomly and uniformly generated $\alpha$ and $\vec{m}$. On one CPU, the DNN model is 3.4 times faster than the analytical model for 1000 cells. However, as the number of cells increases to 10 million, the speedup gain is only 1.3. This is expected, as the overhead associated with moving the tensors to RAM becomes substantial for large batch sizes. The DNN model accelerates significantly better on a GPU (even taking into account the overhead of moving data to the GPU), and with 10 million cells we achieve more than 41 times speedup. In our previous work, we demonstrated that, while traditional algorithms such as Scardovelli and Zaleski \cite{SCARDOVELLI2000228} require fewer floating operations than a neural network approach, the neural network model better utilizes the hardware and operates at higher floating operations per second \cite{ataei2021nplic}.


\subsection{Implementation in a CFD Solver}

LBfoam \cite{ATAEI2021107698} is a free surface lattice Boltzmann solver for foaming simulations. In LBfoam, a disjoining pressure $\Pi$ between two adjacent bubble interfaces is responsible for stabilizing liquid lamellae (the thin layer of liquid separating two bubbles). The disjoining pressure is active up to a distance $\delta_{max}$, and is a function of the distance between bubble interfaces $\delta$ as:

\begin{equation}
	\Pi= \left\{\begin{array}{ll}
        0 & \delta>\delta_{max}\\
        k_{\Pi}\big(1 - \frac{\delta}{\delta_{max}}\big) & \delta<\delta_{max}\\
	\end{array}\right.
	\label{eqn:disjoining}
\end{equation}

\noindent where $k_{\Pi}$ is a constant. In order to calculate $\delta$, LBfoam reconstructs adjacent bubble interfaces with PLIC. 

To verify that the DNN model works as expected in a CFD solver, we replaced the PLIC calculation model in LBfoam with the trained DNN model with $H=2$ and $N=48$.

\begin{figure*}[ht]
	\centering
	\includegraphics[width=0.9\linewidth]{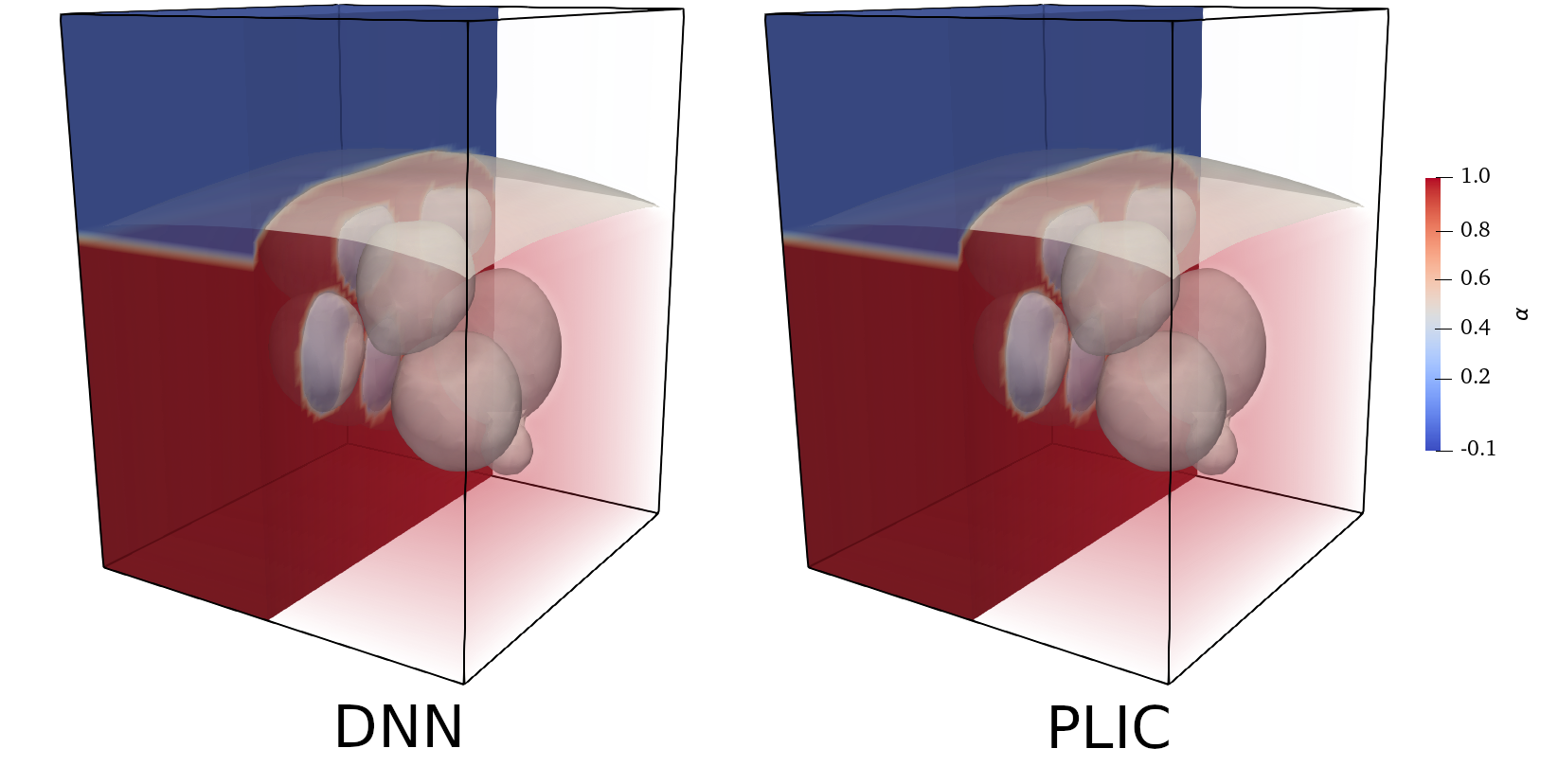}
	\caption{3D foaming simulation using the LBfoam software. The disjoining pressure between bubbles is calculated using the DNN (left) and the analytical PLIC (right).}
	\label{fig:FoamDNN}
\end{figure*}

Figure \ref{fig:FoamDNN} shows simulation results based on the same parameters and at the same timestep. The DNN was used for the simulation on the left, and the analytical PLIC was used on the right. As can be seen, the simulation results are identical. Depending on the number of cells that require interface reconstruction (i.e.\ the interface cells closer than $\delta_{max}$), the speedup gain for interface construction was between 1.5-3 times.

\section{Conclusions}

In this paper, we present a DNN model to solve the forward problem of PLIC. We show that a DNN can learn to solve the PLIC problem without a synthetic dataset, using its inverse problem. The DNN implementation of PLIC is up to several orders of magnitude faster than analytical approaches, for the same accuracy. The DNN implementation is easily portable to various CFD solvers, and can be efficiently accelerated by CPUs and GPUs in parallel. We implemented the DNN model in the LBfoam CFD solver to verify that the model can perform well in a practical setting.

\section*{Acknowledgements}

We thank the Natural Sciences and Engineering Research Council of Canada (NSERC) and Autodesk Inc.\ for their financial support.

\bibliography{main}
\end{document}